\documentclass[epj,nopacs]{svjour}
\usepackage{graphics}
\usepackage{amsmath,amssymb,graphicx}
\usepackage{upgreek}
\begin{document}
\title{Loading an Optical Trap with Diamond Nanocrystals Containing Nitrogen-Vacancy Centers from a Surface}

\titlerunning{Loading from a surface}

\author{Jen-Feng Hsu\inst{1}, Peng Ji\inst{1}, M. V. Gurudev Dutt\inst{1} \and Brian R. D'Urso\inst{1}
}                     
\authorrunning{Hsu et al.}

%
%
\institute{Department of Physics and Astronomy, University of Pittsburgh, Pittsburgh, PA 15260, USA}
%
\date{Received: date / Revised version: date}
%
\abstract{
We present a simple and effective method of loading particles into an optical trap in air at atmospheric pressure. Material which is highly absorptive at the trapping laser wavelength, such as tartrazine dye, is used as media to attach photoluminescent diamond nanocrystals. The mix is burnt into a cloud of air-borne particles as the material is swept near the trapping laser focus on a glass slide. Particles are then trapped with the laser used for burning or transferred to a second laser trap at a different wavelength. Evidence of successfully loading diamond nanocrystals into the trap presented includes high sensitivity of the photoluminecscence (PL) to an excitation laser at 520~nm wavelength and the PL spectra of the optically trapped particles. This method provides a convenient technique for the study of the nitrogen-vacancy (NV) centers contained in optically trapped diamond nanocrystals.
\PACS{
      {PACS-key}{discribing text of that key}   \and
      {PACS-key}{discribing text of that key}
     } 
} 
\maketitle
\section{Introduction}
\label{intro}
The study of nitrogen-vacancy (NV) centers contained in an optically trapped diamond nanocrystal has drawn much experimental interest as a solid-state spin system with an extraordinary long coherence time \cite{Balasubramanian2009Ultralong}. Several characterization results have been reported. For example, electron-spin resonance (ESR) has been observed in both an ensemble \cite{Horowitz2012} and a single spin \cite{geiselmann2013} optically trapped in liquid. NV photoluminescence (PL) has also been observed in air \cite{Neukirch2013}.

Since an optical trap provides a well isolated environment from thermalization of the substrate, feedback cooling of the trapped particle to sub-Kelvin in vacuum can be achieved without cryogenic precooling \cite{Gieseler2012}. This enables the possibility of tests of quantum mechanics such as quantum measurements between spin and mechanical degrees of freedom \cite{DDDN}. Further fundamental experiments including preparing Fock states of the mechanical modes and spatial superposition states have also been proposed \cite{Li2013}.

Various mechanisms have been discovered for forming optical traps, including the electric field gradient force, using either single or dual beams \cite{Ashkin1976,Ashkin1977,Ashkin1978}, a scanned beam \cite{Sasaki1992}, or photophoresis \cite{Jovanovic2009,Nicola1988,Shvedov2010,Zhang2012} and bottle beam configurations \cite{Cacciapuoti2001,Shvedov2011,Chremmos2011,Pan2012,Turpin2013Optical} for absorptive particles.

Multiple methods for loading micron- or nanometer-sized particles into an optical trap have been reported, for example, blowing with air near the trap \cite{Pan2012}, vaporizing the liquid suspension \cite{Neukirch2013}, and ultrasonically shaking particles off a surface \cite{Li2012Fundamental}. One concern of these approaches is that the large volume of air-borne particles might contaminate the optical elements and the experiment chamber.

In this Letter we demonstrate a method of loading diamond nanocrystals into an optical trap from a dye-coated surface in air. The dye solution, which is strongly absorptive at the laser's wavelength, is mixed with diamond nanocrystals and deposited on a glass microscope slide. The proof of the presence of a photoluminescent diamond crystal is the very high sensitivity of PL counts to a 520~nm excitation laser. The exact mechanism for maintaining the particles in the optical trap in our experiment is not completely understood. We observe that absorptive particles are more easily trapped and that the numerical aperture (NA) is very low, as described below, compared with the reported gradient force trapping experiments, e.g., \cite{Omori1997,Gieseler2012,Neukirch2013}. Together with these observations and a further analysis of the spatial distribution of the trapped particle, photophoretic force is speculated to dominate the trapping mechanism.  The detail of the analysis can be found in the Appendix. This method can potentially be applied to many photophoretic-force based experiments, e.g., \cite{Shvedov2010,Shvedov2011,Zhang2012,Pan2012,Turpin2013Optical}. More importantly, we believe that experiments especially involving ensembles of NV centers, e.g., ESR \cite{Horowitz2012} and thermal sensing \cite{Acosta2010Temperature}, can benefit from this loading technique.

\section{Experiment Setup}

\begin{figure}[h!]
\centerline{\includegraphics[width=.75\columnwidth]{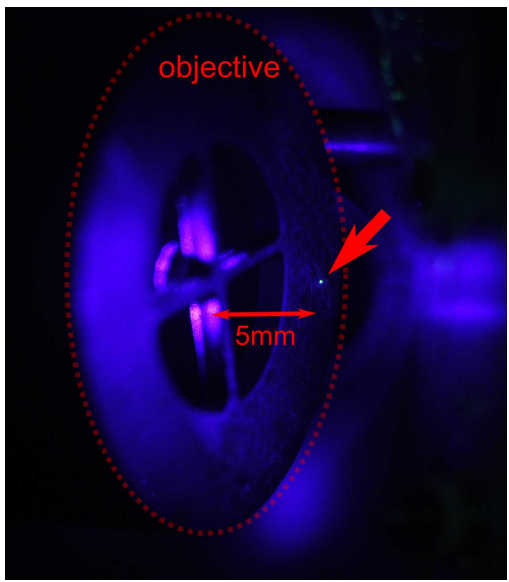}}
\caption{Photograph of the objective and a trapped cluster (indicated by the arrow) illuminated by the excitation laser.}
\label{fig:trapped_particle}
\end{figure}

\begin{figure}[h!]
\centerline{\includegraphics[width=\columnwidth]{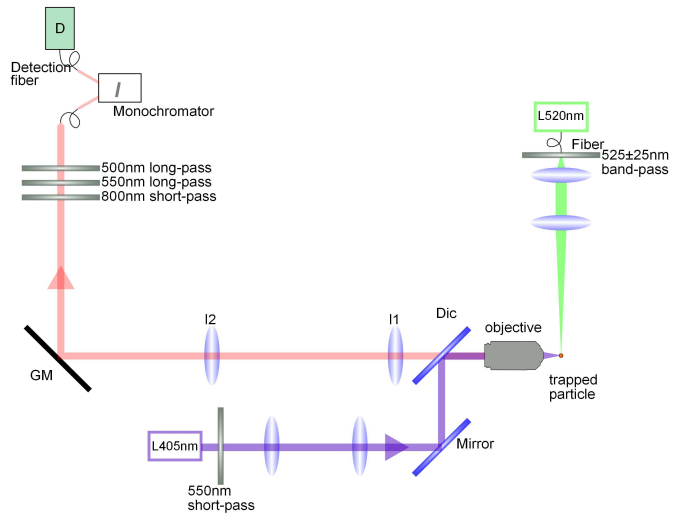}}
\caption{Schematic of the optical setup for this experiment. In the figure, D represents the single photon counting module (SPCM), l$_{1,2}$ the relay lenses, GM the galvo mirror, L405nm and L520nm the laser diodes, and Dic the dichroic mirror.}
\label{fig:optical_setup}
\end{figure}

The optical trap is sourced from coherent light of 405~nm wavelength from a laser diode (the trapping laser). The light from the diode goes through beam-expanding lenses to fill the back of the objective, and then is focused by a reflecting objective with 0.5 NA (Newport 50102-02). The objective is housed in a vacuum chamber. The laser power for loading particles off the glass slide is typically around 50~mW at the exit of the objective. After loading, the power is lowered to about 30~mW, which is roughly minimal to maintain the trap, for characterizing the trapped particle. The 520~nm excitation laser is directed to the optical trap from the side, with a very low NA of 0.07. See Fig.~\ref{fig:trapped_particle} for the objective and a typical trapped cluster illuminated by the excitation laser. With a typical excitation laser power of 0.6~mW, the power density is approximately $\sim$~0.06~mW/$\upmu$m$^2$. In contrast, the trapping laser at 30~mW after the 0.5-NA objective creates a power density of $\sim$~200~mW/$\upmu$m$^2$, several thousand times higher than the excitation laser. Therefore, without a photoluminescent material strongly absorbing at 520~nm, little change in PL when the 520~nm illumination is added should be expected. With a series of edge-pass filters in front of the detector, the PL observed is restricted to the band 550-800~nm. See Fig.~\ref{fig:optical_setup} for the complete experimental optics setup. The monochromator in the setup is used only when recording the PL spectrum. For spatial scans the monochromator is omitted.

	The dye-diamond mixed material to be loaded is prepared by mixing nanodiamonds with an aqueous solution of dye on a glass slide. As the slide is swiped past the focus of the trapping laser, the highly absorptive dye is burnt off from the glass slide and generates a small cloud of air-borne particles. A clump of mixed particles passing by the trap location is then trapped. In this experiment the slurry of diamond nanocrystals of average 100~nm diameter as purchased from Ad\'{a}mas Nanotechnologies (ND-NV-100nm) is used. Tartrazine, the major ingredient of the food coloring Yellow No.~5, is dark orange in color and is strongly absorptive at the wavelength of the trapping laser \cite{Bagirova2003}. Therefore, in this work, tartrazine (Acros AC191891000) is ideal as the absorptive material. Typically 5~$\upmu$L of approximately 0.05~M tartrazine in water solution is mixed with 25~$\upmu$L of the as-purchased nanodiamond solution to form a tartrazine-nanodiamond mix (T-ND). This mix is allowed to dry on a clean microscope glass slide in a circular area of diameter about 5-7~mm.


	As a control experiment, tartrazine dye alone is also prepared on a glass slide and loaded in the same way as T-ND. 100 tartrazine clusters and 100 T-ND clusters were loaded into the trap, one at a time. Loading statistics of both types of clusters are summarized. For loading tartrazine clusters, on average about one out of every ten swipes gives a trapped particle, and about 25\% of them remain trapped long enough for characterization. For loading T-ND clusters, a slightly larger fraction of swipes yields a trapped particle, and about half of them stay long enough for characterization.
		

\section{Results}
\begin{figure}[h!]
\centerline{\includegraphics[width=\columnwidth]{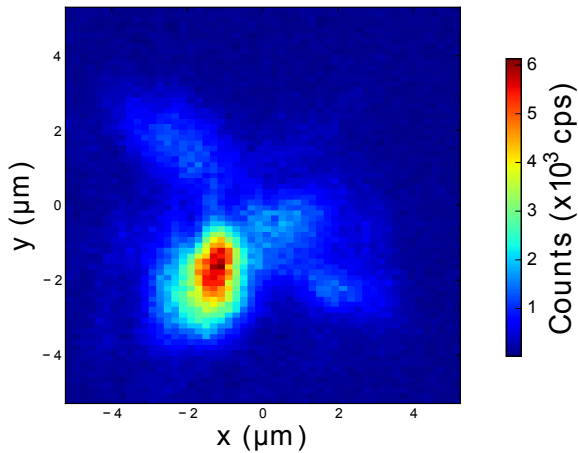}}
\caption{Spatial scan with a single-mode detection fiber of a T-ND cluster with filter edges at 550~nm and 800~nm.}
\label{fig:spatial_scan}
\end{figure}

Figure~\ref{fig:spatial_scan} shows a typical spatial PL scan of a T-ND cluster. With a single-mode fiber, finer structures can be seen and the trapped particle size can be roughly estimated to be 1~$\upmu$m.

\begin{figure}[h!]
\centerline{\includegraphics[width=\columnwidth]{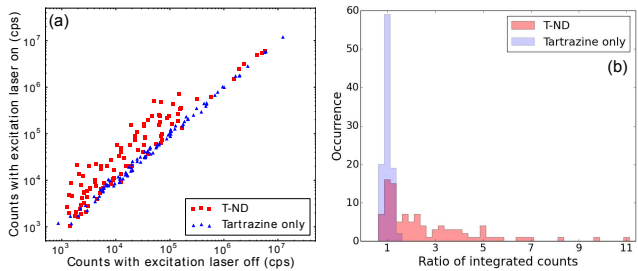}}
\caption{(a) Scattered distribution of integrated counts with and without excitation laser. (b) Histogram of ratio of integrated count with excitation to without excitation laser.}
\label{fig:statistics}
\end{figure}

In each loading, integrated photon counts between 550 and 800~nm with the excitation laser on and off are recorded. Fig.~\ref{fig:statistics}(a) is a scattered plot of the integrated counts. Each point represents one loading and the count with the excitation laser off is the $x$-coordinate and that with the excitation laser on is the $y$-coordinate. From this plot, there is no substantial difference in the distribution of the total counts, as the counts in both cases range from 10$^3$ to 10$^7$~per second. A survey of nanodiamonds scattered on a solid surface also indicates that the PL counts range over three orders of magnitude. This could be caused by a combination of factors including the distribution of the numbers of NVs and their lifetime or orientation in individual nanodiamonds. Clearly from this figure, photon counts by pure tartrazine clusters are tightly distributed around the $y=x$ line. This indicates that the counts are not dependent on whether the excitation laser is on or off. However, at least 50\% of the clusters from T-ND have higher counts by several times when the excitation laser is on. This signifies a much higher sensitivity of T-ND to the excitation laser.

To quantitatively see the difference in the counts with or without the excitation laser, Fig.~\ref{fig:statistics}(b) shows the histogram of the ratios of the counts with the excitation laser to that without the excitation laser. For pure tartrazine clusters, the ratios are tightly centered around one, with just few occurrences approaching 1.5. The ratios of counts for T-ND span a range from 1 to 11 and for about 25\% of the particles the ratios take on values larger than 3. This test indicates that the material loaded from the T-ND mixture is much more photoluminescent, while nanodiamonds are the only different material that were added. It shall be stressed again that the power densities of the trapping and excitation lasers are differed by 3 orders of magnitude. Taking this into account, the histogram $x$-axis could be scaled by a factor of 1000 to plot the ratio of sensitivity to the trap and excitation lasers.

\begin{figure}[h!]
\centerline{\includegraphics[width=\columnwidth]{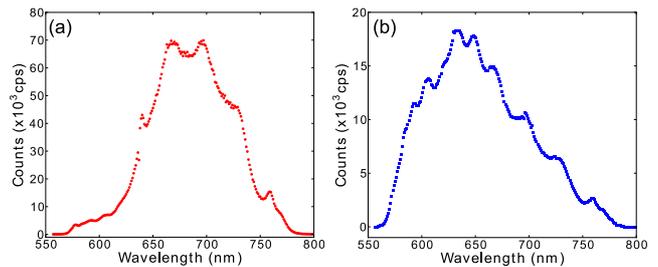}}
\caption{PL spectra of (a) nanodiamond and (b) tartrazine on a Si wafer surface. Excitation power 40 $\upmu$W. Filter edges are at 550~nm and 800~nm.}
\label{fig:spectra_on_Si}
\end{figure}

Next we study the PL spectra of the two different materials. With a monochromator (resolution approximately 1.2~nm), as a reference we first record the spectra of nanodiamond and tartrazine on a Si wafer, as shown in Fig.~\ref{fig:spectra_on_Si}(a) and \ref{fig:spectra_on_Si}(b), respectively. Here the trapping laser is kept off.

In Fig.~\ref{fig:spectra_on_Si}(a), peaks at 578 and 640~nm are discernible. These peaks are qualitatively consistent with the zero-phonon lines (ZPL) shown in the manufacturer's data of the nanodiamond. For tartrazine only as in Fig.~\ref{fig:spectra_on_Si}(b), the spectrum has some different structure, but the ZPL of nanodiamonds is not seen and the overall band shape is shifted to the shorter wavelengths. These spectra will be used to compare with that of the optically trapped particles.

\begin{figure}[h!]
\centerline{\includegraphics[width=\columnwidth]{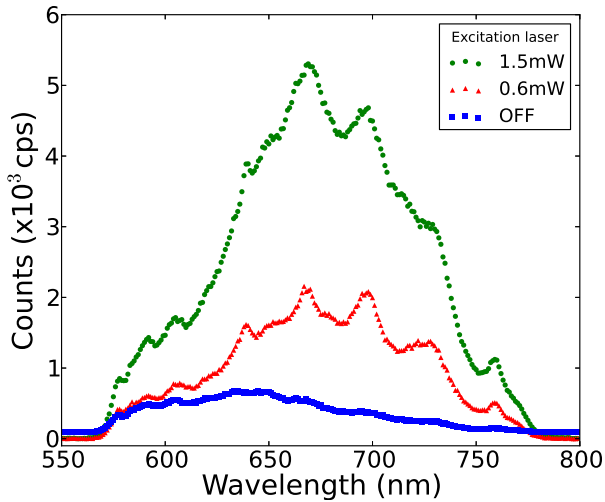}}
\caption{PL spectra of trapped T-ND with various excitation laser powers. Filter edges are at 550~nm and 800~nm.}
\label{fig:T-ND_spectrum_in_trap}
\end{figure}

Subsequently, the spectral structures of optically trapped T-ND and tartrazine are studied. In Fig.~\ref{fig:T-ND_spectrum_in_trap}, the spectra of the optically trapped T-ND with different excitation laser powers are shown. Prominent sensitivity to the excitation laser is present, consistent with the data in Fig.~\ref{fig:statistics}. Also, the ZPL lines and the general band shape are consistent with the spectrum of nanodiamonds on a Si surface as well. Note the red-shifted band when the excitation laser is turned on. 

\begin{figure}[h!]
\centerline{\includegraphics[width=\columnwidth]{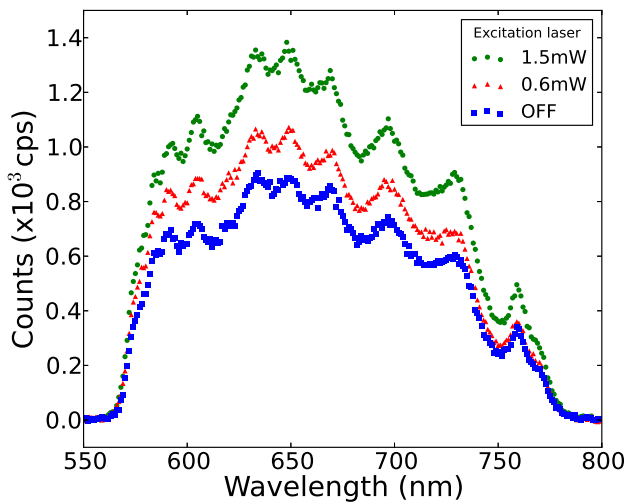}}
\caption{PL spectra of trapped tartrazine with various excitation laser powers. Filter edges are at 550~nm and 800~nm.}
\label{fig:tartrazine_spectrum_in_trap}
\end{figure}

In Fig.~\ref{fig:tartrazine_spectrum_in_trap}, the spectral shape of a tartrazine cluster in an optical trap is similar with the that on a Si surface. With the same excitation powers as in Fig.~\ref{fig:T-ND_spectrum_in_trap} for T-ND, a much weaker dependence of the spectrum on the excitation power is present.

\section{Discussion}
\label{sec:discussion}

It is observed from Fig.~\ref{fig:T-ND_spectrum_in_trap} that the excitation power is not yet enough to saturate the NVs in the trapped T-ND cluster. However, for excitation power higher than what is shown in Fig.~\ref{fig:T-ND_spectrum_in_trap}, we have also observed that the excitation laser radiation force tends to drive the particle out of the optical trap.

Because even the pure tartrazine clusters can stay in the optical trap for many hours (long enough for characterization before the trapping laser is turned off), the tartrazine is apparently not completely burnt away by the trapping laser. Instead, we speculate that the sodium content is burnt into sodium oxide, which is stable at high temperature. The significance of this possibility is that for the trapped T-ND, the material left in the trap can be nanodiamond plus other compounds, which may affect the characterization of the nanodiamonds. 

In addition to the presence of the 405~nm trapping laser, oxidation of tartrazine could also account for the difference in the spectral shapes of tartrazine on a Si surface and that in an optical trap, as in Fig.~\ref{fig:spectra_on_Si}(b) and Fig.~\ref{fig:tartrazine_spectrum_in_trap}, respectively. The tartrazine on a Si surface was not illuminated by the 405~nm trapping laser and the 520~nm excitation laser is not enough to oxidize the tartrazine. So, it is likely that the trapped tartrazine is chemically different from the unoxidized tartrazine on a Si surface, and thus has a different spectrum.

In the effort to minimize the possible residue still attached to the nanodiamonds after loading, we have also observed loading of nanodiamonds from a surface coated with phenol red-dyed cellulose nitrate flim (NC). This combination is employed to optimize the clean burning of material since phenol red does not contain sodium and NC burns cleanly. Further utilizing the explosive nature of NC, the surface could efficiently release nanodiamonds from the surface by burning minimal amount of the material with a focused laser. However, we found that the loading efficiency is still lower than that with tartrazine. We speculate that this is due to the cleanliness of the nanodiamond surface, which potentially makes the photophoretic force much weaker. See Appendix for discussion on the relative strengths of the trapping forces.


\section{Conclusion}
	We demonstrate a new method of loading photoluminescent diamond nanocrystals into an optical trap. This method avoids large number of suspended particles, which potentially contaminate the optics and the chamber. Using an absorptive dye material as media for burning and loading, this method is effective and the rate of successfully loading photoluminescent material is very high. PL spatial and spectral scans are recorded. Very high sensitivity to the excitation laser is observed. Both the sensitivity and the spectral shape of the optically trapped particles are strong evidence of loading nanodiamonds containing NV centers.	The diamond can simply be much cleaner.


\section*{Acknowledgment}
	The authors thank Chenxu Liu of University of Pittsburgh for help with part of the NV characterization in this experiment. G. D. acknowledges support by NSF CAREER (DMR-0847195), NSF PHY-1005341, DOE Early Career (DE-SC0006638), and the Alfred P. Sloan Research Fellowship.


\section*{Appendix: Analysis of the Trapping Mechanism}
\label{trapping_width}

\begin{figure}[h!]
\centerline{\includegraphics[width=0.8\columnwidth]{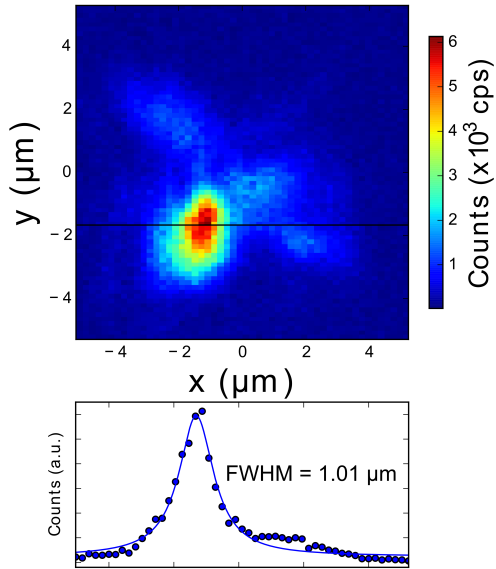}}
\caption{Spatial scan (top) and profile distribution (bottom) with a single-mode detection fiber of a T-ND cluster with filter edges at 550~nm and 800~nm. The profile is taken along the black line on the spatial scan.}
\label{fig:spatial_scan_line_profile}
\end{figure}

Our optics setup allows us the nearly full resolution of the objective. The full diameter of the back opening of the objective can be imaged onto the mode field diameter of the single-mode detection fiber without clipping. Hence the resolution of the spatial scan shown in Fig.~\ref{fig:spatial_scan} and Fig.~\ref{fig:spatial_scan_line_profile} is limited by the focused beam diameter, which is roughly 400~nm for the trapping light and 650~nm for the PL. 

As shown in Fig.~\ref{fig:spatial_scan_line_profile}, the trapped particle is distributed in a fitted width of approximately 1~$\upmu$m. In comparison, the spatial distribution is larger than the beam diameter. More importantly, in Fig.~\ref{fig:spatial_scan_line_profile}, the spatial structure that spans over $\pm$2~$\upmu$m in both directions seems to indicate that the trapped particle moves between a few different trap locations. Based on these observations, the trapping mechanism in this experiment is believed to be the photophoretic force.

Incidentally, though it is observed in our experiment that trapped particles are lost while pumping down in pressure, this is not a conclusive evidence for the photophoretic nature of the trapping. A gradient-force-trapped silica bead also tends to escape the trap while pumping down if without feedback cooling \cite{Li2012Fundamental}. The thermal kinetic energy might already be large enough to overcome the trapping potential. 

Finally, order-of-magnitude estimation of photophoretic, gradient, and scattering forces for diamond nanospheres in an optical trap are compared. Photophoretic force can be calculated by the formulae given in \cite{Wurm2010Experiments} with the thermal conductivity of diamond, which is typically between 1000 and 3000~W/m/K. This results in a force at low pico-Newton order. The scattering force for our system falls in the same range with photophoretic force. The gradient force can be estimated similarly as in \cite{Gieseler2012} with the NA and wavelength of our setup and the relative permittivity of diamond, typically between 5 to 8. This leads to a force at the 10-pico-Newton order. This may seem surprising at a first look. It does not seem to be consistent with other published results that photophoretic forces are orders of magnitudes stronger, e.g., \cite{Shvedov2009Optical}, and our experimental observations also suggest that photophoretic force is dominant. We suggest the following explanation to resolve these contradictions. First, the photophoretic force is inversely proportional to the thermal conductivity of the material \cite{Wurm2010Experiments}, which spans a wide range of many orders of magnitude. Due to diamond's well-know superior thermal conductivity, the photophoretic force on a clean diamond particle can be orders of magnitude weaker than that on absorbing materials, such as soot, burnt dye, or multi-walled carbon nanotubes \cite{Pan2012}. Second, this calculation suggests that our trapped particles, though mainly of diamond, are likely covered by a layer of burnt dye, which is dark and absorbing. Photophoretic force can therefore be much stronger. This can also explain the lower loading efficiency with NC, as mentioned in Sec.~\ref{sec:discussion}. The diamond particle coming off NC is probably simply cleaner.

%

%

\end{document}